\documentclass[aps,prl,reprint,superscriptaddress]{revtex4-1}
\pdfoutput=1
\usepackage{amsmath,amssymb}	% ams math packages
\usepackage{bm}				% bold math symbols
\usepackage{color}				% color
\usepackage{graphicx}			% graphics
\begin{document}

\title{Physiologically motivated multiplex Kuramoto model describes phase diagram of cortical activity}

\author{Maximilian Sadilek}
%\email[]{maximilian.sadilek@meduniwien.ac.at}
%\homepage[]{http://www.complex-systems.meduniwien.ac.at}
%\thanks{}
\affiliation{Section for Science of Complex Systems, Medical University of Vienna, Spitalgasse 23, A-1090, Vienna, Austria}
\author{Stefan Thurner}
\email[]{stefan.thurner@meduniwien.ac.at}
%\homepage[]{http://www.complex-systems.meduniwien.ac.at}
%\thanks{Corresponding author}
\affiliation{Section for Science of Complex Systems, Medical University of Vienna, Spitalgasse 23, A-1090, Vienna, Austria}
\affiliation{Santa Fe Institute, 1399 Hyde Park Road, New Mexico 87501, USA}
\affiliation{IIASA, Schlossplatz 1, A-2361 Laxenburg; Austria}

%\date{\today}

\begin{abstract}
We derive a two-layer multiplex Kuramoto model from weakly coupled Wilson-Cowan oscillators on a cortical network with inhibitory synaptic time delays. Depending on the coupling strength and a phase shift parameter, related to cerebral blood flow and GABA concentration, respectively, we numerically identify three macroscopic phases: unsynchronized, synchronized, and chaotic dynamics. These correspond to physiological background-, epileptic seizure-, and resting-state cortical activity, respectively. We also observe frequency suppression at the transition from resting-state to seizure activity.
\end{abstract}

\maketitle

%%%%%%%%%%%%%%%%%%%%%%%%%%%%%%%%%%%%%%%%%%%%%%%%%%%%%%%%%%%%%%%%%%%
%\section{\label{sec:introduction}Introduction}
%%%%%%%%%%%%%%%%%%%%%%%%%%%%%%%%%%%%%%%%%%%%%%%%%%%%%%%%%%%%%%%%%%%
Fast electrochemical processes taking place on a complicated cytoarchitectural network structure render the human brain a highly complex dynamical system. 
Brain activity, as measured directly via EEG or MEG, or indirectly by means of MRI recordings, reveals characteristic macroscopic patterns such as oscillations in various frequency bands~\cite{osci}, synchronization~\cite{epilepsy,schizo,brainweb}, or chaotic dynamics~\cite{eeg}.
Generally it is believed that macroscopic activity (involving $10^8-10^{11}$ neurons) is closely related to high-level functions such as cognition, attention, memory or task execution. 
To understand the mechanisms of this correspondence, both the overall network structure of the brain and the local properties of neural populations have to be taken into account~\cite{brainweb,cbn}.
Regarding the latter, neural inhibition seems to be essential for cortical processing~\cite{inhib}.

Two physiological phenomena received much attention lately in terms of a mathematical understanding: resting-state activity, and gamma oscillations.
Resting-state activity is spontaneous, highly structured activity of the brain during rest,
and can be described in terms of networks of simultaneously active brain regions~\cite{rsn,deco}.
Models of resting-state networks often rely on anatomical networks derived from histological or imaging data, and 
on local interactions between populations of excitatory and inhibitory neurons~\cite{honey,ghosh,deco2,rs}.
Oscillatory neural activity in the gamma range ($30-100$ Hz) is potentially related to consciousness and the 
binding problem although its precise function remains unclear~\cite{gray}.
To understand the origin of gamma oscillations, two mechanisms have been proposed~\cite{gamma}.
One describes interactions between inhibitory neurons together with an external driving force~\cite{whittington,ii}. 
The other mechanism is based on excitatory-inhibitory coupling with synaptic time delays \cite{brunel,geisler,wc}. 
The relation of gamma oscillations and inhibition is experimentally well established. 
In mice~\cite{mice}, rats~\cite{rats} and humans~\cite{gaba,schizo}, a decrease of GABA-concentrations
(Gamma-aminobutyric acid is the main inhibitory neurotransmitter in mammals) is accompanied by a strong 
attenuation of the gamma frequency band and sometimes by epileptiform activity. 
%%%%%%%%%%%%%%%%%%%%%%%%%%%%%%%%%%%%%%%%%%%%%%%%%%%%%%%%%%%%%%%%%%%
\begin{figure*}[t]
\includegraphics[width=0.33\textwidth]{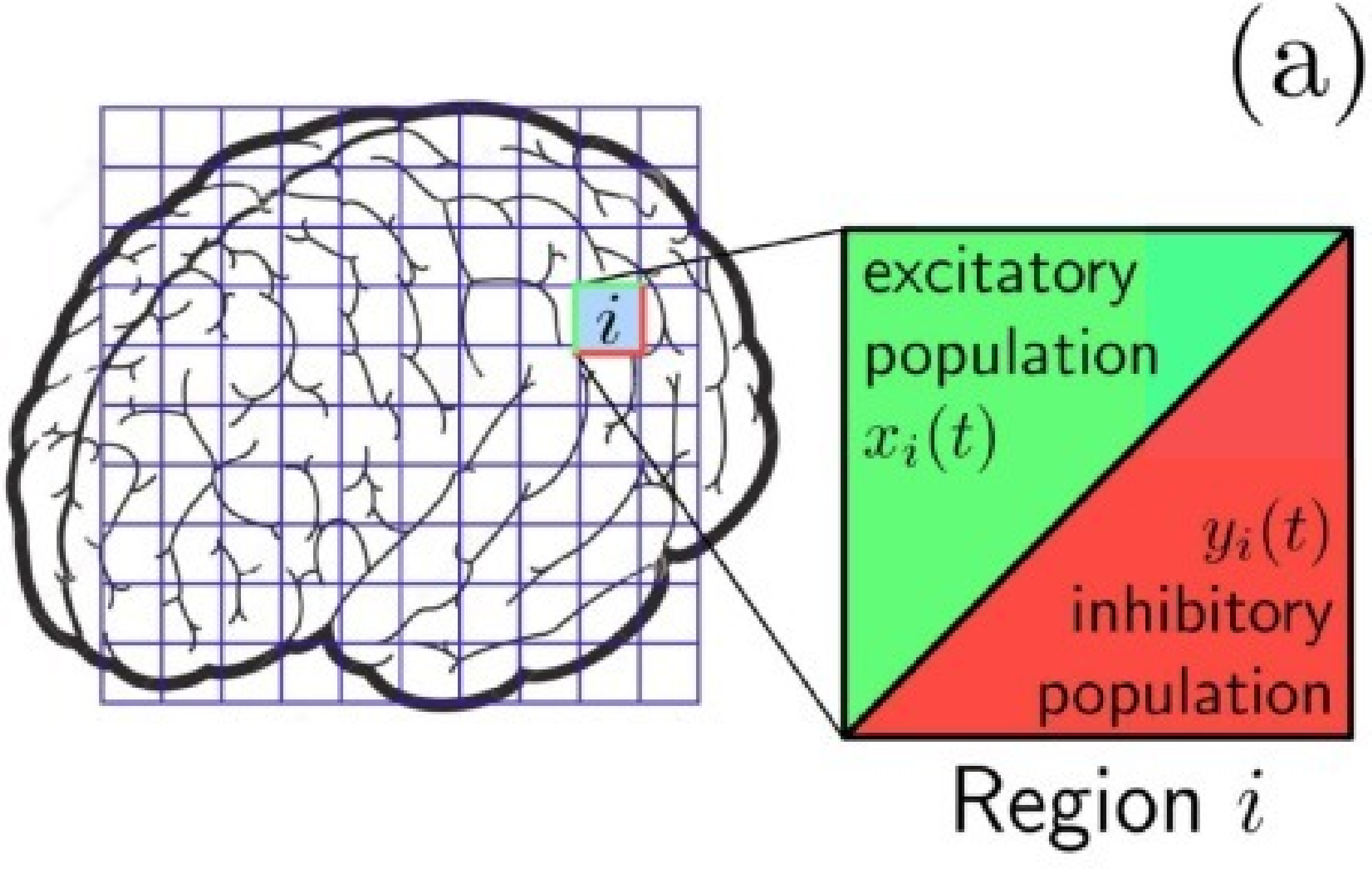} 
\includegraphics[width=0.33\textwidth]{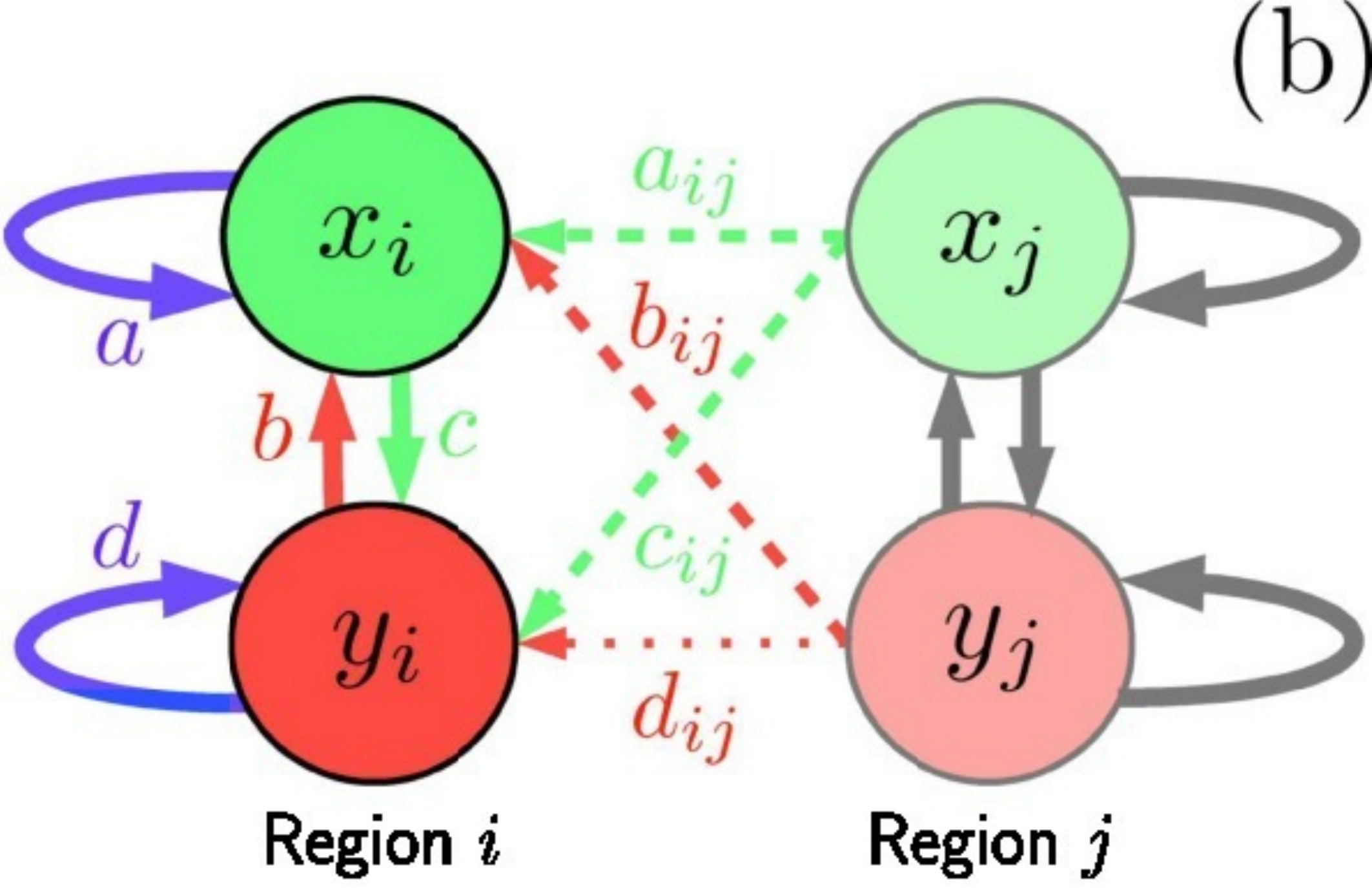} 
\includegraphics[width=0.329\textwidth]{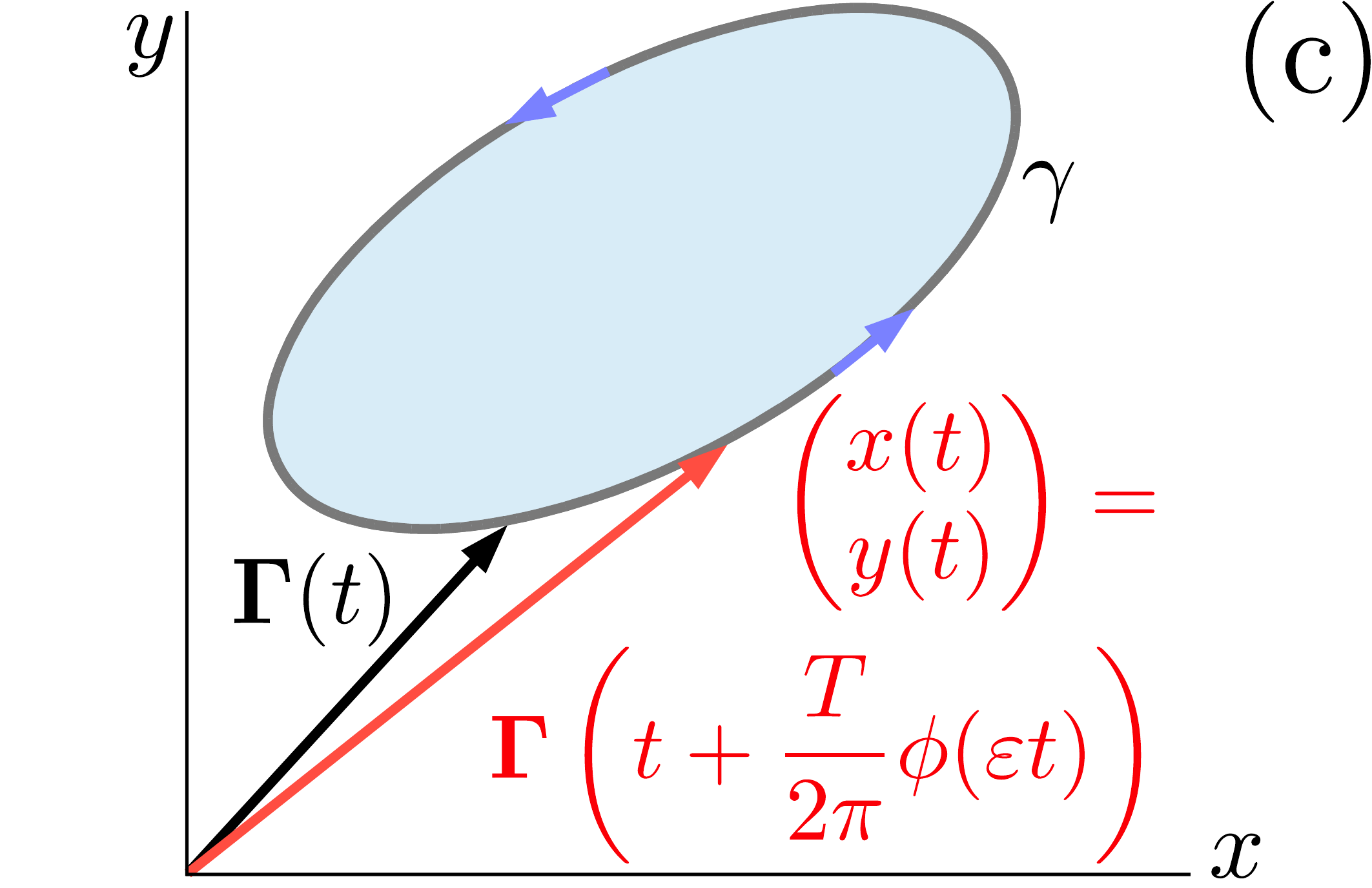}
\caption{(a) The cortical surface is divided into $N$ macroscopic regions. 
Every region $i$ (blue) comprises excitatory (green) and inhibitory (red) populations with activity levels $x_i(t)$ and $y_i(t)$, respectively. 
Activity levels quantify the ratio of firing neurons in the region at time $t$. 
(b) Region $i$ (left) receives excitatory (green arrows) and inhibitory (red arrows) inputs from adjacent regions $j$ (right) plus self-feedback (blue arrows). 
(c) Variable transformation from activity variables $x_i(t)$ and $y_i(t)$ to phase deviation variables $\phi_i(\varepsilon t)$. 
On a limit cycle $\gamma$, $\varepsilon$-perturbations of the $(x,y)$-dynamics at time $t$ induce the phase deviations $\phi(\varepsilon t)$.}
\label{fig:1}
\end{figure*}
%%%%%%%%%%%%%%%%%%%%%%%%%%%%%%%%%%%%%%%%%%%%%%%%%%%%%%%%%%%%%%%%%%%

Many existing network models for resting-state activity and gamma oscillations are based on single-neuron local dynamics~\cite{honey,ghosh,whittington,ii,brunel,geisler}.
Since experimentally observed resting-state networks comprise individual regions containing about $10^8$ to $10^9$ individual neurons, we believe that a local description in terms of Wilson-Cowan equations is an attractive alternative.
In this work we derive a simple two-layer multiplex model from classical physiological equations that is 
able to capture the main features of cortical activity such as oscillations, synchronization and chaotic dynamics. 
This model unifies the roles of neural network topology, synaptic time delays, and excitation/inhibition. 
It provides a closed framework for simultaneously understanding the origin of resting-state activity and gamma oscillations. 
%%%%%%%%%%%%%%%%%%%%%%%%%%%%%%%%%%%%%%%%%%%%%%%%%%%%%%%%%%%%%%%%%%%
%\section{\label{sec:model}Model}
%%%%%%%%%%%%%%%%%%%%%%%%%%%%%%%%%%%%%%%%%%%%%%%%%%%%%%%%%%%%%%%%%%%

We consider $N$ cortical regions indexed by $i=1,\dots N$, see Fig.~\ref{fig:1}(a). 
Each region is populated by ensembles of excitatory and inhibitory neurons (e.g.\ pyramidal cells and interneurons). 
We define the activity level of a region $i$ as the fraction of firing excitatory (inhibitory) neurons of the total number of excitatory (inhibitory) neurons in that region at a unit time interval, and denote it by $x_i$ ($y_i$).
Neglecting for the moment interactions between different regions, we assume that individual cortical regions obey the Wilson-Cowan-type dynamics \cite{wc}
\begin{align}\label{eq:wc}
\dot{x_i}&=-x_i+S\left(a_i x_i -b_i y_i+\rho^{(x)}_i\right)\\
\dot{y_i}&=-y_i+S\left(c_i x_i -d_i y_i+\rho^{(y)}_i\right)\quad,\nonumber
\end{align}
where 
$S$ is a sigmoidal response function, 
$a_i$ and $d_i$ are real-valued feedback parameters, and $b_i$ and $c_i$ are positive synaptic coefficients ($i=1\dots N$). 
$\rho^{(x)}_i$ and $\rho^{(y)}_i$ account for external inputs, e.g.\ from sensory organs. 
We now introduce interactions among regions of the network by replacing $\rho_i^{(\cdot)}$ in Eq.~(\ref{eq:wc}) by
\begin{align}\label{eq:ans}
\rho^{(x)}_i\rightarrow\rho^{(x)}_i&+\sum_{j\neq i}\left[a_{ij}x_j(t)- b_{ij}y_j(t-\tau)\right]\\
\rho^{(y)}_i\rightarrow\rho^{(y)}_i&+\sum_{j\neq i}\left[c_{ij}x_j(t-\tau)-d_{ij}y_j(t)\right]\quad.\nonumber
\end{align}
Here $a_{ij}$, $b_{ij}$, $c_{ij}$, $d_{ij}$ are positive synaptic coefficients linking regions $i$ and $j$. 
$\tau$ accounts for transmission delays at inhibitory synapses (not to be confused with axonal conduction delays). 
In physiology, $\tau$ can be altered by changing the synaptic concentration of GABA~\cite{brunel,geisler,mice,rats}. 
In the present model, we assume that $\tau$ is proportional to the average synaptic GABA concentration in the brain.
To derive a multiplex Kuramoto model (MKM) from Eqs. (\ref{eq:wc}) and (\ref{eq:ans}), we make the following three assumptions:

%\begin{enumerate}[(i)]
%\item 
{\em (i)  Homogeneity.} Cortical regions exhibit nearly identical dynamical behavior. 
We therefore assume the following parameters to be constant across regions, 
\begin{equation}\label{eq:params}
\quad\quad(a_i,b_i,c_i,d_i,\rho_i^{(x)},\rho_i^{(y)})=(a,b,c,d,\rho^{(x)},\rho^{(y)})+\mathcal{O}(\varepsilon)
\end{equation}
for all $i$, up to small perturbations, denoted by  $\varepsilon$.

%\item 
{\em (ii) Stable local oscillations.} We choose the parameters $(a,b,c,d,\rho^{(x)},\rho^{(y)})$ such that each uncoupled system 
Eq.~(\ref{eq:wc}), under the assumption given in Eq.~(\ref{eq:params}), has a unique exponentially stable limit cycle 
$\gamma\in\mathbb{R}^2$. As a consequence, after a transient time solutions of Eq.~(\ref{eq:wc}) can be written as
\begin{equation}\label{eq:gammadef}
\begin{pmatrix}
x_i(t)\\y_i(t)
\end{pmatrix}
=\bm{\Gamma}(t+\varphi_i)\quad,
\end{equation}
where $\bm{\Gamma}(t)$ is an arbitrary solution of Eq.~(\ref{eq:wc}) on $\gamma$. $\varphi_i$ accounts for specific initial values~\cite{bk,izh}. 
Let $T$ denote the period of $\bm{\Gamma}$.
We assume that the frequency $1/T$ lies in the physiological gamma range.

%\item 
{\em (iii) Weak coupling.} Interactions between adjacent regions are weak, and inhibitory-inhibitory interactions are very weak in the sense that,
\begin{equation}\label{eq:weak}
(a_{ij},b_{ij},c_{ij},d_{ij})=\varepsilon(A_{ij},B_{ij},C_{ij},0)+\mathcal{O}(\varepsilon^2)\quad,
\end{equation}
for all $i\neq j$. 
These assumptions are justified because the number of synaptic connections within a cortical region is much larger than between regions,
and excitatory neurons outnumber inhibitory neurons by approximately one order of magnitude~\cite{fairen,defelipe}. 
%\end{enumerate}

Figure~\ref{fig:1}(b) summarizes the connectivity structure between regions $i$ and $j$. 
Region $i$ receives excitatory (green arrows) and inhibitory (red arrows) inputs plus feedback (blue arrows), magnitudes are indicated by the arrow labels. 
For the sake of clarity, arrows representing inputs of magnitude $\mathcal{O}(1)$, $\mathcal{O}(\varepsilon)$ and $\mathcal{O}(\varepsilon^2)$ are drawn in continuous, dashed and dotted style, respectively.

Under these assumptions, the  system Eq.~(\ref{eq:wc}) with Eq.~(\ref{eq:ans}) is equivalent~\cite{supp} to a two-layer MKM 
\begin{align}\label{eq:model}
\frac{d\phi_i}{dt}&=\omega_i+\frac{K}{\left<k\right>}\sum_{j=1}^NA_{ij}\sin(\phi_j-\phi_i)\\
&+\frac{K}{\left<k^{(\delta)}\right>}\sum_{j=1}^{N}A^{(\delta)}_{ij}\sin(\phi_j-\phi_i-\delta)\quad.\nonumber 
\end{align}
Here $\phi_i(t)\in \mathcal{S}^1$ describes the deviations from the uncoupled phases 
$\theta_i(t)\equiv2\pi/T (t+\varphi_i)$ that are associated with solutions of the uncoupled system Eq.~(\ref{eq:gammadef}). 
Accordingly, $d\phi_i/dt$ describes the deviations from the uncoupled oscillation frequency $2\pi/T$.
Time $t$ has been rescaled~\cite{supp}. 
$A_{ij}$ is the adjacency matrix of the excitatory-excitatory interaction network as defined in Eq.~(\ref{eq:weak}), 
$A_{ij}^{(\delta)}\equiv1/2\left(B_{ij}+C_{ij}\right)$ accounts for the interaction between excitatory and inhibitory populations, and $\left<k\right>\equiv1/N\sum_{i,j}^NA_{ij}$, and $\left<k^{(\delta)}\right>\equiv1/N\sum_{i,j}^NA_{ij}^{(\delta)}$, are the corresponding average degrees.

$\delta$ is a phase shift parameter related to the time delay $\tau$ via $\delta\equiv(2\pi/T)\;\tau$.
$K$ is a global coupling constant that we assume to be proportional to the cerebral blood flow. 
This is reasonable because the latter is strongly correlated with the connection strengths of functional networks reconstructed in magnetic resonance imaging~\cite{dfmri}. 
$\omega_i$, the so-called \emph{natural frequencies} of the MKM, are the constant contribution to the frequency deviations $d\phi_i/dt$.
We take $\omega_i$ from a symmetric, unimodal random distribution $g(\omega)$, with mean $\omega_0$.
Since the 1-parameter family of rotating-frame transformations $\phi_i(t)\rightarrow \phi_i(t)-\Omega t$, $\omega_i\rightarrow \omega_i-\Omega$, leave Eq.~(\ref{eq:model}) invariant for any $\Omega$, without loss of generality we assume, $\omega_0=0$ and $\delta\in[0,\pi]$.
Note that for each solution $\phi_i(t)^*$ with $\delta^*\in[\pi,2\pi]$, there exists a solution $\phi_i(t)$ with $\delta=2\pi-\delta^*\in[0,\pi]$ and $\phi_i(t)=-\phi_i(t)^*$. 
Physiological processes changing $\delta$, $K$, and $\omega_i$ occur on a much slower  timescale than neural activity.

It is known that weakly coupled, nearly identical limit-cycle oscillators can be described in terms of phase variables~\cite{winfree,kura,kura2,izh}.
However, in terms of the new variables $(\phi_1\dots \phi_N)$, interactions between cortical regions take place on two independent 
layers representing excitatory-excitatory and excitatory-inhibitory coupling, respectively, and 
the complicated connectivity structure of Fig.~\ref{fig:1}(b) reduces to a simple two-layer multiplex structure.
Figure~\ref{fig:1}(c) shows the variable transformation from activity variables $x$ and $y$, to the phase variable $\phi$ for any cortical region. 
In the unperturbed case, $\varepsilon=0$, the limit cycle $\gamma$ is parametrized by $\bm{\Gamma}(t)=(x(t),y(t))$. 
Since $\gamma$ is exponentially stable, $\varepsilon$-perturbations of activity dynamics $t\mapsto \left(x(t),y(t)\right)$ lead to phase deviations $\bm{\Gamma}(t+T/(2\pi) \phi(\varepsilon t))$.
For $\delta=0$ we recover the Kuramoto model on a single network, see~\cite{supp,arenas,kalloniatis,llekura,yeung,enzo,niebur}.

\emph{Synchronization}. We define the order parameter~\cite{kura,kura2,acebron,arenas}
\begin{equation}\label{eq:r}
r(t)\equiv \left| \frac{1}{N} \sum_{j=1}^{N} \exp \left[i \phi_j(t)\right] \right|\quad.
\end{equation}
It takes values between $0$ (no synchronization) and $1$ (full synchronization)~\cite{izh}.
Let $\overline r$ denote its time average $\overline r\equiv\left<r(t)\right>_t$.

\emph{Chaotic dynamics}. The instantaneous largest Lyapunov exponent is given by
\begin{equation}\label{eq:lle}
\lambda_{max}(t) \equiv \lim_{d_0\to 0}\frac{1}{t-t_0}\ln\frac{d(t)}{d_0}\quad,
\end{equation}
where $d(t)=\left\Vert \phi(t)-\phi^p(t)\right\Vert$ measures the separation between a reference trajectory $\phi(t)$ and a perturbed one $\phi^p(t)$. 
$d_0$ is the initial separation at $t_0$, and $\left\Vert\cdot\right\Vert$ is the $1$-norm~\cite{supp}.
For large times, $\lambda_{max}(t)$ approaches the ``true" largest Lyapunov exponent, $\overline\lambda_{max}$. 

\emph{Average frequency deviation}. We look at average frequency deviations across all regions,
\begin{equation}\label{eq:omega}
\Omega\equiv\frac{1}{N}\sum_{i=1}^N\frac{d\phi_i}{dt}\quad,
\end{equation}
once a stationary state is reached.\\
%%%%%%%%%%%%%%%%%%%%%%%%%%%%%%%%%%%%%%%%%%%%%%%%%%%%%%%%%%%%%%%%%%%
\begin{figure*}
\centering
\includegraphics[width=0.24\textwidth]{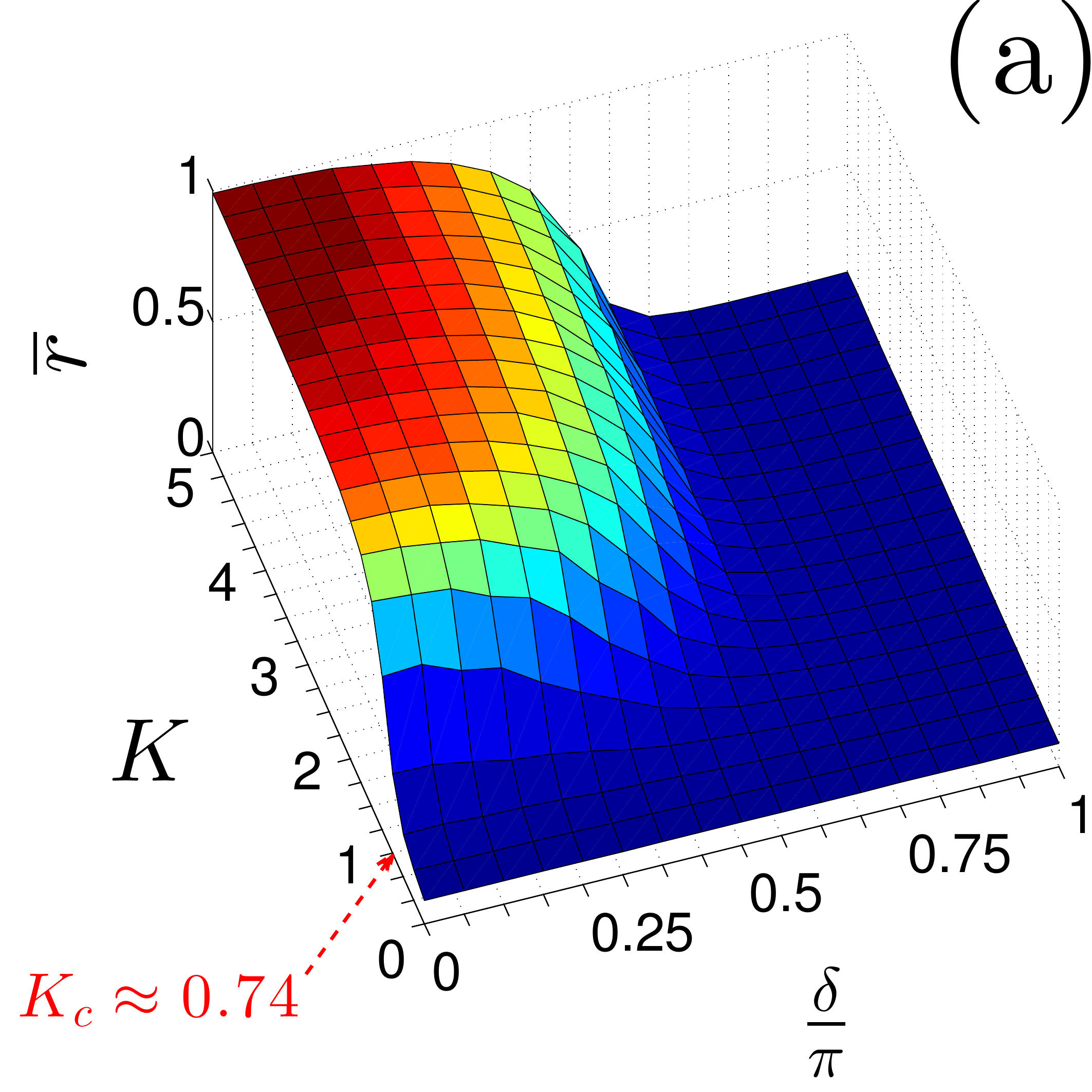}
\includegraphics[width=0.24\textwidth]{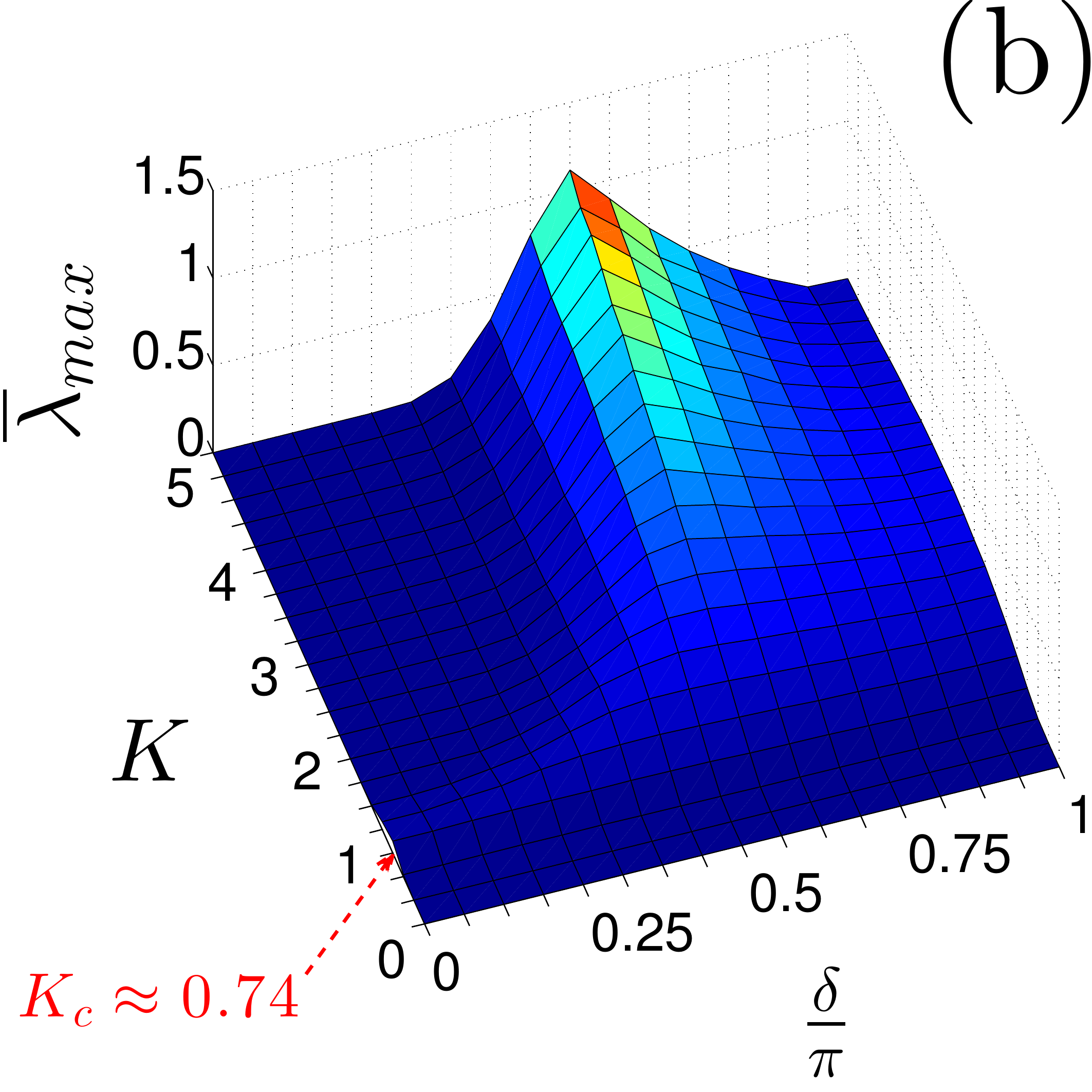}
\includegraphics[width=0.24\textwidth]{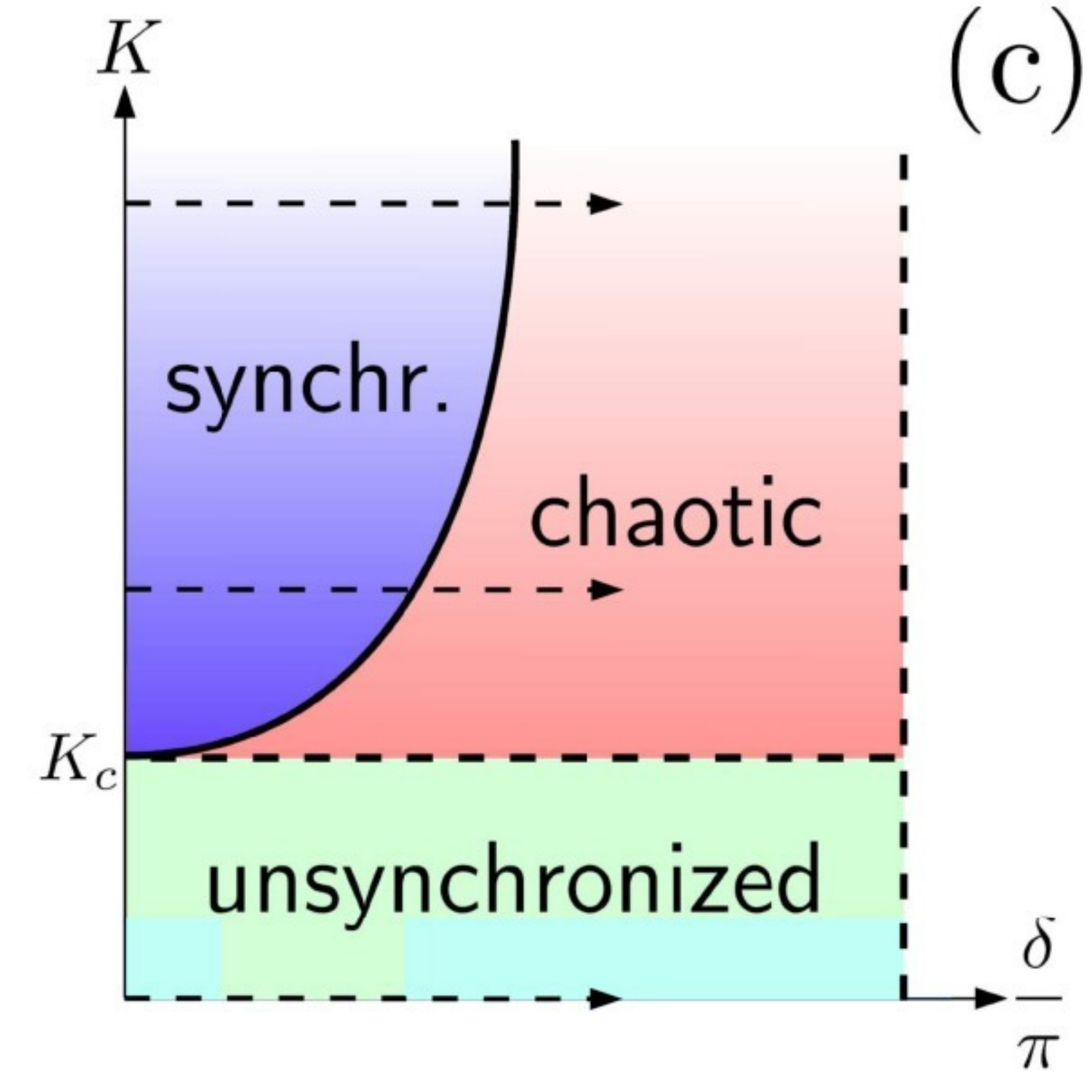}
\includegraphics[width=0.24\textwidth]{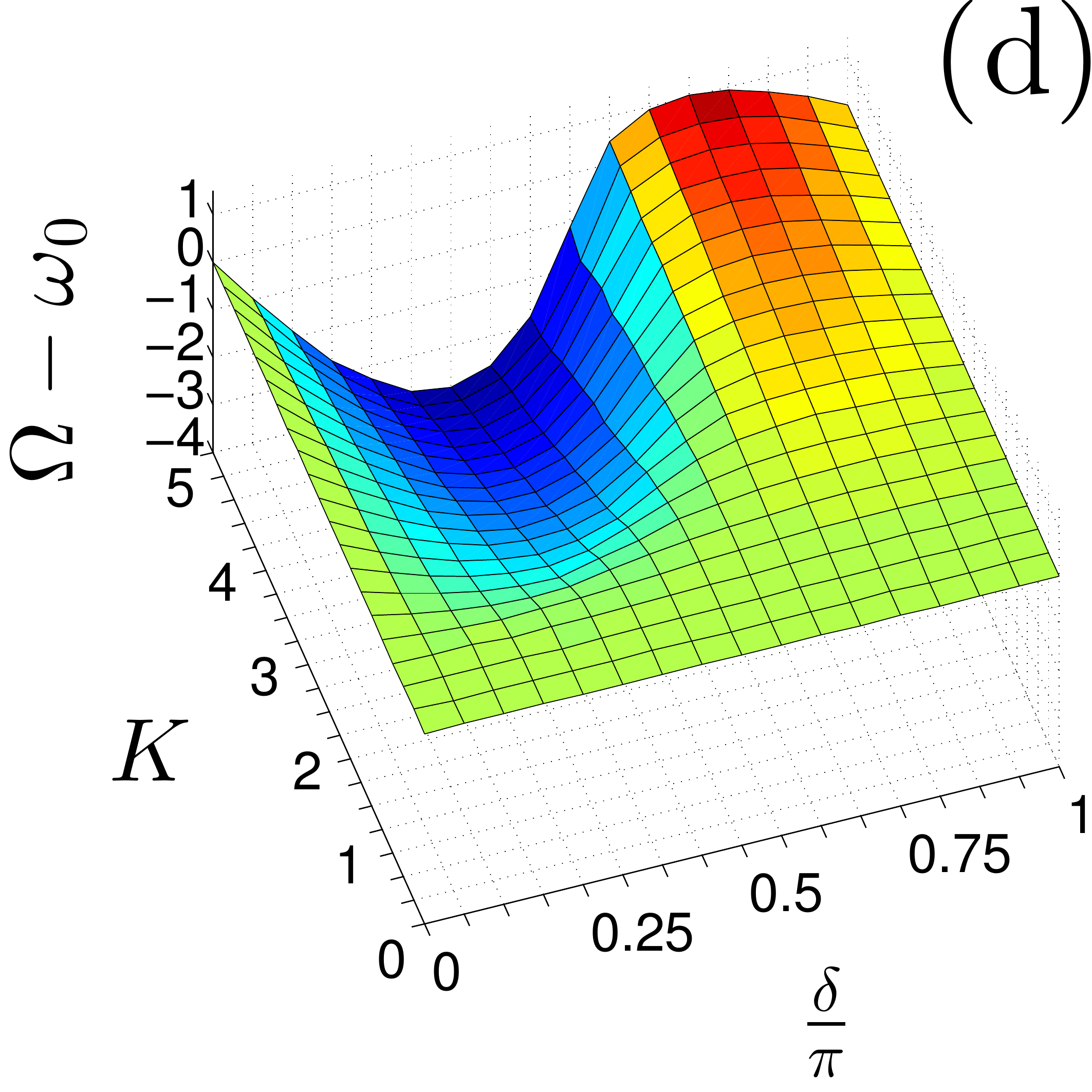} 
\caption{(a) Order parameter $\overline r$ and (b) largest Lyapunov exponent $\overline\lambda_{max}$ identify the regions of synchronization and of chaotic dynamics in the $(K,\delta)$-plane.
(c) Schematic phase diagram.
Dashed arrows indicate directions along which distributions of frequency deviations were evaluated in Figs.~\protect{\ref{fig:3}}(a)-(c).
(d) Average frequency deviations $\Omega$ in the $(K,\delta)$-plane.}
\label{fig:2}
\centering
\end{figure*}
%%%%%%%%%%%%%%%%%%%%%%%%%%%%%%%%%%%%%%%%%%%%%%%%%%%%%%%%%%%%%%%%%%%
%\section{\label{sec:results}Results}
%%%%%%%%%%%%%%%%%%%%%%%%%%%%%%%%%%%%%%%%%%%%%%%%%%%%%%%%%%%%%%%%%%%

Equation~(\ref{eq:model}) is integrated with a standard 4th-order Runge-Kutta algorithm with $500$ time steps of size $dt=0.1$. 
The system size is $N=100$, both layers are chosen to be Erd\H{o}s-R\'{e}nyi networks with $p=0.06$. 
Natural frequencies $\omega_i$ are taken from a standard normal distribution, initial phase deviations $\phi_i(0)$ from the interval $[0,2\pi]$. 
The first $250$ time steps are discarded to exclude transient effects.
For the remaining time steps, $\overline r$, $\overline\lambda_{max}$, and $\Omega$ are evaluated.
All results are averaged over $100$ identical, independent runs with different realizations of the initial conditions.

\textit{Synchronization}.
We find that synchronization $\overline r$ depends on the coupling strength $K$, and phase shift $\delta$, Fig.~\ref{fig:2}(a).
For $\delta=0$, we expect \cite{supp} a transition from an unsynchronized to a synchronized state at a critical value $K_c\approx0.74$, which is confirmed by our simulations, Fig.~\ref{fig:2}(a).
With $\delta>0$, stronger coupling $K$ is required for this transition to occur. 
Above a value of approximately $\delta=\pi/2$, a global synchronized state ceases to exist.

\textit{Chaotic dynamics}.
Above the synchronization threshold, $K>K_c$, synchronization and chaotic dynamics are mutually exclusive, see Fig.~\ref{fig:2}(b).
For small values of $\delta$, there exists a small chaotic region ($\overline\lambda_{max}\approx 0.1$) at the Kuramoto transition, in agreement with the well-known results for $\delta=0$,~\cite{supp}.
This region is expanding with increasing values of $\delta$.
At the boundary to the synchronized region, increasingly large values of $\overline\lambda_{max}$ are obtained.
$\overline\lambda_{max}$ peaks at $\overline\lambda_{max}\approx 1$, for $\delta \approx \pi/2$.
In the unsynchronized region, $K<K_c$, the dynamics is not chaotic, $\overline\lambda_{max}=0$.
Figure~\ref{fig:2}(c) integrates both results (synchronization and chaotic dynamics) into a schematic 
phase diagram that clearly exhibits three phases. 

\begin{figure}
\includegraphics[width=0.47\textwidth]{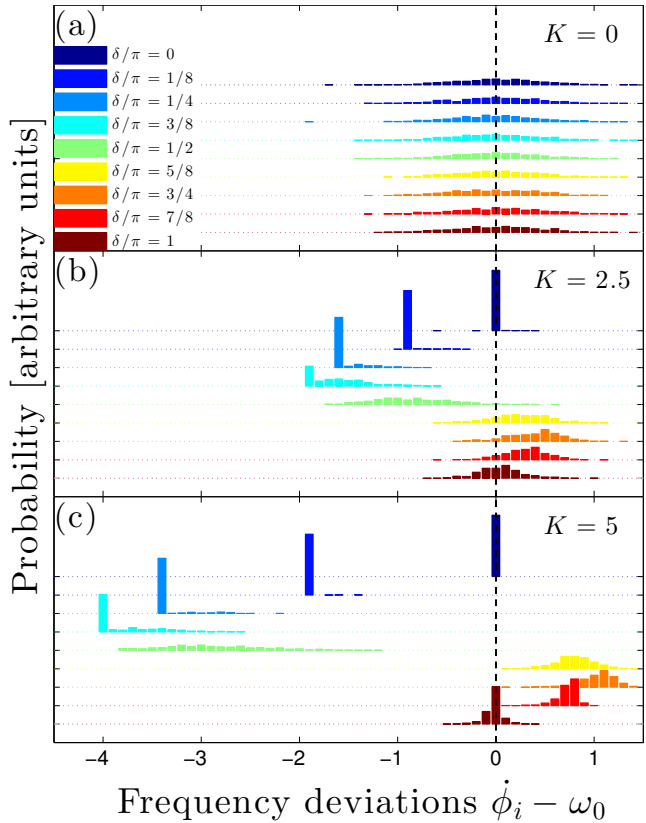}
\caption{Stationary distribution of frequency deviations $d\phi_i/dt$ for different $\delta$ values (colors) for
(a) subcritical, (b) critical and (c) supercritical values of $K$, respectively. $K_c\approx 0.74$.}
\label{fig:3}
\end{figure}

\textit{Spectral properties}.
Figure~\ref{fig:3} (a)-(c) shows the stationary distributions of frequency deviations $d\phi_i/dt$ for selected values in the ($K$,$\delta$)-plane. 
For $K=0$, the distributions are practically identical for different values of $\delta$, Fig.~\ref{fig:3}(a). 
For $K=2.5$, at $\delta=0$, a synchronization peak appears close to frequency zero. 
With increasing $\delta$, this peak moves towards increasingly negative values, until $\delta\approx3\pi/8$. 
Between $\delta\approx3\pi/8$ and $3\pi/4$, the distribution is rapidly becoming broader and shifts towards positive values. 
After reaching a maximum at $\delta\approx3\pi/4$, it is finally centered around zero again, Fig.~\ref{fig:3}(b). 
$K=5$, is similar, however larger positive and negative values for $d\phi_i/dt$ occur, Fig.~\ref{fig:3}(c). 

Figure~\ref{fig:2}(d) shows the average frequency deviation $\Omega$ as a function of $K$ and $\delta$. 
As expected~\cite{supp}, we find frequency suppression associated with synchronization in the region of large $K$ and small $\delta$, but also for large $K$ and intermediate $\delta$. 
For fixed $K$, maximal frequency suppression occurs at $\delta\approx3\pi/8$. 
For large $K$ and large $\delta$ (chaotic regime) we find slightly positive $\Omega$.
%%%%%%%%%%%%%%%%%%%%%%%%%%%%%%%%%%%%%%%%%%%%%%%%%%%%%%%%%%%%%%%%%%%
%%%%%%%%%%%%%%%%%%%%%%%%%%%%%%%%%%%%%%%%%%%%%%%%%%%%%%%%%%%%%%%%%%%
%\section{\label{sec:Discussion}Discussion}
%%%%%%%%%%%%%%%%%%%%%%%%%%%%%%%%%%%%%%%%%%%%%%%%%%%%%%%%%%%%%%%%%%%

In several variants of single-layer Kuramoto models with a phase shift or a time delay, frequency suppression appears~\cite{niebur,howto,choi,enzo}. In addition, ~\cite{enzo} mentions chaotic behavior.
However, the existence of the phase diagram with the three distinct macroscopic phases can not be inferred from any of those models to the best of our knowledge.

We tested the model for robustness with respect to the particular choice of parameters. 
As suggested by various brain atlases and cortical parcellation schemes, a number of $N\approx100 - 150$ cortical regions seems reasonable~\cite{brod,parc}. 
We tested up to $N=500$ and found no deviations from the presented qualitative picture. 
For the link density $p$, we find that as long as it exceeds the percolation threshold, $p>\log(N)/N$, differences in simulations are marginal. 
Finally, we observe that like in the original Kuramoto model~\cite{kura,kura2}, for different natural frequency distributions $g(\omega)$ the qualitative behaviour remains practically unchanged as long as $g(\omega)$ is unimodal and symmetric.

Summarizing, we can show mathematically that a set of weakly coupled Wilson-Cowan oscillators on a cortical network with a synaptic time delay between excitatory and inhibitory neural populations is identical to a simple Kuramoto-type phase model on a two-layer multiplex network. Numerical investigations of this model reveal the presence of three distinct macroscopic phases in the space of control parameters $K$ (associated with cerebral blood flow) and $\delta$ (associated with synaptic GABA concentration).
For couplings $K<K_c$, activities of individual cortical regions show independent oscillatory behavior (unsynchronized). 
Frequencies are distributed symmetrically around an average frequency that we assume to be located in the physiological gamma range. 
This dynamical state corresponds to ``background activity'' of the brain.
For $K>K_c$, two phases are possible: for small $\delta$, the system becomes synchronized, which corresponds to ``epileptic seizure activity'' in physiology. 
For large $\delta$, synchronized activity only appears in metastable clusters; the system is chaotic in general. 
We identify this phase with ``resting-state activity" in the brain.
An important property of the present model is that the average oscillation frequency is shifted towards lower values when crossing the boundary to the synchronized phase.
This could explain the experimental fact~\cite{epilepsy,gaba,mice,rats} that a decrease of the GABA concentration in the resting-state both triggers the appearance of epileptiform slow waves and diminishes gamma activity in the brain.

We acknowledge financial support from EC FP7 projects LASAGNE, agreement no. 318132, and MULTIPLEX, agreement no. 317532.
%%%%%%%%%%%%%%%%%%%%%%%%%%%%%%%%%%%%%%%%%%%%%%%%%%%%%%%%%%%%%%%%%%%
%\pagenumbering{gobble}

%%%%%%%%%%%%%%%%%%%%%%%%%%%%%%%%%%%%%%%%%%%%%%%%%%%%%%%%%%%%%%%%%%%
\newpage
%\pagebreak
\begin{widetext}
\begin{center}
\textbf{\large Supplemental Material}
\end{center}

%%%%%%%%%% Prefix a "S" to all equations, figures, tables and reset the counter %%%%%%%%%%
%\setcounter{equation}{0}
%\setcounter{figure}{0}
%\setcounter{page}{1}
%\renewcommand{\thepage}{SM\arabic{page}}
\makeatletter
\renewcommand{\theequation}{S\arabic{equation}}
%\renewcommand{\bibnumfmt}[1]{[S#1]}
%\renewcommand{\citenumfont}[1]{S#1}
%%%%%%%%%% Prefix a "S" to all equations, figures, tables and reset the counter %%%%%%%%%%

\newcommand{\T}{\frac{T}{2\pi}}

%%%%%%%%%%%%%%%%%%%%%%%%%%%%%%%%%%%%%%%%%%%%%%%%%%%%%%%%%%%%%%%%%%%
\section{Derivation of the multiplex Kuramoto model}

In this section, we derive the MKM, Eq.~(6), from nearly identical, 
weakly coupled Wilson-Cowan oscillators, see Eqs.~(1) - %,eq:ans,eq:params,eq:gammadef,
(5).
The main idea is the following: Starting from uncoupled oscillators having exponentially stable orbits, the introduction of weak coupling between them only affects their phases (and not their frequencies or amplitudes). This allows for a transformation from activity- to phase deviation variables. The special structure of the Wilson-Cowan equations then accounts for the multilayer structure of the MKM.

For compact notation, we define ${\bm{X}}_i\equiv (x_i,y_i)^{\top}$ and ${\bm{X}}\equiv ({\bm{X}}_1,\dots {\bm{X}}_N)$, where $^{\top}$ denotes transposition. 
Assumptions expressed in Eqs.~(3) and (5), allow us to write Eq.~(1) with the interactions introduced by Eq.~(2) in the form
\begin{equation}\label{eq:full}
\frac{d{\bm{X}}_i(t)}{dt}={\bm{F}}\left[{\bm{X}}_i(t)\right]+\varepsilon\,\delta {\bm{F}}_i\left[{\bm{X}}_i(t)\right]+\varepsilon\,{\bm{G}_i}\left[{\bm{X}}(t),{\bm{X}}(t-\tau)\right]\quad,
\end{equation}
up to terms of order $\mathcal{O}(\varepsilon^2)$. 
Here ${\bm{F}}$ constitutes the uncoupled part of the dynamics, and $\delta {\bm{F}}_i$ accounts for small perturbations in the parameters $a,b,c,d,\rho^{(x)}$, and $\rho^{(y)}$. 
${\bm{G}_i}$ encodes the contributions from weak coupling. 
Terms of order $\mathcal{O}(\varepsilon^2)$ will be omitted in all subsequent expressions.

Assumption from Eq.~(4) states that solutions ${\bm{X}}_i^*(t)$ of the uncoupled system 
\begin{equation*}
\frac{d{\bm{X}}_i(t)}{dt}={\bm{F}}\left[{\bm{X}}_i(t)\right]
\end{equation*}
approach a unique, exponentially stable limit cycle $\gamma\subset \mathbb{R}^{2}$. 
This means that after a transient phase we can write ${\bm{X}}_i^*(t)=\bm{\Gamma}(t+\varphi_i)$. 
We define the phase of such a solution by the mapping $\theta\colon {\bm{X}}_i^*(t)\mapsto 2\pi/T (t+\varphi_i)\in \mathcal{S}^1$, uniquely assigning a point on the unit cycle to each point of the solution, where $T$ is the period of ${\bm{\Gamma}}$.

Since solutions ${\bm{X}}_i^*(t)$ of the uncoupled system are exponentially orbitally stable, we can write solutions of the weakly coupled system Eq.~(\ref{eq:full}) as
\begin{equation}\label{eq:ansatz}
{\bm{X}}_i(t)=\bm{\Gamma}\left[t+\T\phi_i(\varepsilon t)\right]+\varepsilon{\bm{P}_i}\left[t+\T\phi_i(\varepsilon t)\right]\quad,
\end{equation}
where $\phi_i(t)\in\mathcal{S}^1$ is the deviation from the phase $\theta_i(t)\equiv\theta({\bm{X}}_i^*(t))=2\pi/T (t+\varphi_i)$, see Fig.~1(c). 
$\varepsilon {\bm{P}_i}$ accounts for the effects of the $\varepsilon$-perturbation of the invariant manifold $\gamma^N\subset \mathbb{R}^{2N}$, where $\gamma^N$ denotes the $N$-th Cartesian power of $\gamma$,~\cite{izh}. 
Initial values of $\phi_i$ are given by
\begin{equation*}
\phi_i(0)=\frac{2\pi}{T}\varphi_i\quad.
\end{equation*}
Differentiating Eq.~(\ref{eq:ansatz}) with respect to $t$ and using Eq.~(\ref{eq:full}), we arrive at the linear inhomogeneous equation
\begin{equation}\label{eq:lin}
\frac{d{\bm{v}_i}(t,\phi_i)}{dt}={\bm{M}(t,\phi_i)}\;{\bm{v}_i}(t,\phi_i)+{\bm{m}_i}(t,\bm{\phi})\quad,
\end{equation}
where $\bm{\phi}\equiv\left(\phi_1(\varepsilon t),\dots \phi_N(\varepsilon t)\right)$ and
\begin{align*}
{\bm{v}_i}(t,\phi_i)&\equiv {\bm{P}_i}\left(t+\T\phi_i\right)\quad,\\
{\bm{M}(t,\phi_i)}&\equiv D{\bm{F}}\left(t+\T\phi_i\right)\quad,\\
{\bm{m}_i}(t,\bm{\phi})&\equiv {\bm{G}_i}\left(t+\T\bm{\phi},t-\tau+\T\bm{\phi}\right)+\delta {\bm{F}}_i\left(t+\T\phi_i\right) - {\bm{F}}\left(t+\T\phi_i\right)\frac{d\phi_i}{d(\varepsilon t)}\quad.
\end{align*}
Here $D{\bm{F}}$ denotes the Jacobian matrix of ${\bm{F}}$, and we have used the abbreviations ${\bm{F}}(t)\equiv {\bm{F}}\left(\bm{\Gamma}(t)\right)$, $\delta {\bm{F}}_i(t)\equiv \delta {\bm{F}}_i\left(\bm{\Gamma}(t)\right)$, and ${\bm{G}_i}(t,t-\tau)\equiv {\bm{G}_i}\left(\bm{\Gamma}(t),\bm{\Gamma}(t-\tau)\right)$ for clarity of notation. 
Now, since both solutions ${\bm{v}_i}(t,\phi_i)$ of Eq.~(\ref{eq:lin}) and ${\bm{w}(t,\phi_i)}$ of the adjoint \emph{homogeneous} system
\begin{equation}\label{eq:adj}
\frac{d{\bm{w}(t,\phi_i)}}{dt}=-{\bm{M}(t,\phi_i)}^{\top}\;{\bm{w}(t,\phi_i)}
\end{equation}
are periodic, nontrivial and unique by construction, the orthogonality condition
\begin{equation}\label{eq:orth}
\frac{1}{T}\int_0^T{\bm{w}(t,\phi_i)}^{\top}{\bm{m}_i}(t,\bm{\phi})\;dt=0\quad,
\end{equation}
follows (Fredholm alternative). 
If we normalize the solution ${\bm{w}(t,\phi_i)}$ of Eq.~(\ref{eq:adj}) by the condition
\begin{equation}\label{eq:norm}
\frac{1}{T}\int_0^T{\bm{w}(t,\phi_i)}^{\top} \;\T {\bm{F}}\left(t+\T\phi_i\right)\;dt=1\quad,
\end{equation}
insert the definition of ${\bm{m}_i}(t,\bm{\phi})$ into Eq.~(\ref{eq:orth}), and use Eq.~(\ref{eq:norm}) we obtain
\begin{equation*}
\frac{d\phi_i}{d(\varepsilon t)}=\frac{1}{T}\int_0^T{\bm{w}(t,\phi_i)}^{\top}\left[\delta {\bm{F}}_i\left(t+\T\phi_i\right)+{\bm{G}_i}\left(t+\T\bm{\phi},t-\tau+\T\bm{\phi}\right)\right]dt\quad.
\end{equation*}
Due to Eq.~(\ref{eq:adj}) and the special form of ${\bm{M}(t,\phi_i)}$ we have ${\bm{w}(t,\phi_i)}={\bm{w}}\left(t+\T\phi_i,0\right)$. 
If we define ${\bm{Q}}(t)\equiv {\bm{w}}(t,0)$ and make the substitutions $t\to t+\T\phi_i$ and $\varepsilon t \to t$, we obtain
\begin{equation}\label{eq:prel}
\frac{d\phi_i}{dt}=\frac{1}{T}\int_0^T{\bm{Q}}(t)^{\top}\left\{\delta {\bm{F}}_i\left[\bm{\Gamma}\left(\vphantom{\T}t\right)\right] + {\bm{G}_i}\left[\bm{\Gamma}\left(t+\T\left(\bm{\phi}-\phi_i\right)\right),\bm{\Gamma}\left(t-\tau+\T\left(\bm{\phi}-\phi_i\right)\right)\right]\right\}\;dt\quad,
\end{equation}
where we expanded the shortened notation for ${\bm{F}}$, $\delta{\bm{F}}_i$ and ${\bm{G}}_i$ again. Making use of the structure of ${\bm{G}_i}$, see 
Eqs.~(1), (2), (\ref{eq:full}), we can rewrite Eq.~(\ref{eq:prel}) as

\begin{equation}\label{eq:heq}
\frac{d\phi_i}{dt}=\omega_i+\sum_{j\neq i}^NA_{ij}H_{EE}(\phi_j-\phi_i)+B_{ij}H_{EI}(\phi_j-\phi_i-\delta)+C_{ij}H_{IE}(\phi_j-\phi_i-\delta)\quad,
\end{equation} 
where $\delta$ is a phase shift parameter related to the time delay $\tau$ through $\delta=(2\pi/T)\tau$, and
\begin{align*}
\omega_i&\equiv\frac{1}{T}\int_0^T{\bm{Q}}(t)^{\top}\delta {\bm{F}}_i\left[\bm{\Gamma}(t)\right]\,dt\\
H_{EE}(\chi)&\equiv\frac{1}{T}\int_0^TQ_1(t)\;S'\hspace{-2pt}\left(ax(t)-by(t)+\rho^{(x)}\right)\;x\hspace{-2pt}\left(t+\T\chi\right)\,dt\\
H_{EI}(\chi)&\equiv-\frac{1}{T}\int_0^TQ_1(t)\;S'\hspace{-2pt}\left(ax(t)-by(t)+\rho^{(x)}\right)\;y\hspace{-2pt}\left(t+\T\chi\right)\,dt\\
H_{IE}(\chi)&\equiv\frac{1}{T}\int_0^TQ_2(t)\;S'\hspace{-2pt}\left(cx(t)-dy(t)+\rho^{(y)}\right)\;x\hspace{-2pt}\left(t+\T\chi\right)\,dt\quad.
\end{align*}
Here $S'(x)\equiv dS/du|_{u=x}$, $Q_1$ and $Q_2$ are the two components of ${\bm{Q}}$, and $x\left(t+\T\chi\right)$ and $y\left(t+\T\chi\right)$ are the two components of $\bm{\Gamma}\left(t+\T\chi\right)$, respectively.
Following the argument of Kuramoto~\cite{kura,kura2}, we keep only the sinusoidal terms of a Fourier series expansion of the right hand side of Eq.~(\ref{eq:heq}), 
and obtain 
\begin{equation}\label{eq:fast}
\frac{d\phi_i}{dt}=\omega_i+\sum_{j\neq i} K_A A_{ij}\sin(\phi_j - \phi_i)+\sum_{j\neq i}K_A^{(\delta)}A_{ij}^{(\delta)}\sin(\phi_j - \phi_i-\delta)\quad,
\end{equation}
where $A_{ij}^{(\delta)}\equiv B_{ij}+C_{ij}$, and $K_A$ and $K_A^{(\delta)}$ are constants. 
Finally, we set
\begin{equation*}
K_A \left<k\right>=K_A^{(\delta)}\left<k^{(\delta)}\right>\equiv K\quad,
\end{equation*}
where $\left<k\right>\equiv 1/N\sum_{i,j}^NA_{ij}$ and $\left<k^{(\delta)}\right>\equiv 1/N\sum_{i,j}^NA_{ij}^{(\delta)}$ are the average degrees of the excitatory-excitatory and excitatory-inhibitory networks, respectively.
This ensures that the two interaction terms on the right hand side of Eq.~(\ref{eq:fast}) are of the same order.
With this notation we obtain the two-layer MKM, i.e. Eq.~(6),
\begin{equation}\label{eq:model2}
\frac{d\phi_i}{dt}=\omega_i+\frac{K}{\left<k\right>}\sum_{j\neq i}^NA_{ij}\sin(\phi_j-\phi_i)
+\frac{K}{\left<k^{(\delta)}\right>}\sum_{j\neq i}^{N}A^{(\delta)}_{ij}\sin(\phi_j-\phi_i-\delta)\quad.
\end{equation}
%
%%%%%%%%%%%%%%%%%%%%%%%%%%%%%%%%%%%%%%%%%%%%%%%%%%%%%%%%%%%%%%%%%%%
\section{The case $\delta=0$}

In this section, we investigate some of the properties of the single-layer Kuramoto model, 
\begin{equation}\label{eq:delta0}
d\phi_i/dt=\omega_i+K/\left<k\right>\sum_{j=1}^N\alpha_{ij}\sin(\phi_j-\phi_i)\quad,
\end{equation}
that is obtained by setting $\delta=0$ in Eq.~(6), and introducing the abbreviation $\alpha_{ij}\equiv A_{ij}+A_{ij}^{(\delta)}\left<k\right>/\left<k^{(\delta)}\right>$.

\emph{Synchronization}.  For sufficiently large $N$ and sufficiently dense networks $A_{ij}$ and $A_{ij}^{(\delta)}$, we expect the following behavior~\cite{arenas,kalloniatis}: 
For small values of $K$, the system attains a stationary global state that shows no synchronization, $\overline r\sim0$, up to finite-size fluctuations of order $\mathcal{O}(1/\sqrt{N})$.
Above a critical value,
\begin{equation}\label{eq:kcrit}
K_c\equiv K_c^0\left<k\right>\frac{\left<\kappa\right>}{\left<\kappa^2\right>}\quad,
\end{equation}
synchronized clusters appear, $\overline r>0$, where $K_c^0\equiv2/\left(\pi g(0)\right)$, is the critical value for a fully connected network, and $\left<\kappa\right>$ and $\left<\kappa^2\right>$ are the first two moments of the distribution of degrees $\kappa_i\equiv\sum_{j=1}^N\alpha_{ij}$, respectively. If both $A_{ij}$ and $A_{ij}^{(\delta)}$ are ErdÒ\H{o}Ós-R\'{e}nyi networks with probability $p$, $\kappa_i$ follows a binomial distribution with probability $2p$. Hence we get $K_c=K_c^0/2\left(1-(1-2p)/2pN\right)+\mathcal{O}\left(N^{-2}\right)$, 
meaning that the Kuramoto transition occurs at slightly lower values than for fully connected networks, as long as $\log(N)/N<p<0.5$.
For $K>>K_c$, global synchronization is attained, $\overline r\sim 1$, after a transient time $\propto \left<k\right>/(K\lambda_2)$, where $\lambda_2$ is the second-smallest eigenvalue of the Laplacian, $L_{ij}\equiv\kappa_i\delta_{ij}-\alpha_{ij}$.

\emph{Chaotic dynamics}. The largest Lyapunov exponent takes the value $\overline\lambda_{max}=0$ in the subcritical regime, and becomes positive around $K=K_c$, reaching peak values of $\overline\lambda_{max}\approx 0.05$. 
For $K>K_c$, $\overline\lambda_{max}$ decreases to marginally negative values $\overline\lambda_{max}\approx -0.001$,~\cite{llekura}.

\emph{Average frequency deviation}.
For large values of $K$, solutions of Eq.~(\ref{eq:delta0}) are globally synchronized, and $\Omega=1/N\sum_{i=1}^N\omega_i\approx \omega_0$. 
Related models~\cite{yeung,niebur,enzo} suggest that small phase shifts $\delta>0$ do not destroy synchronization, so
 we expect $\Omega\approx\omega_0-K/(\left<k^{(\delta)}\right>N)\sum_{i,j}A_{ij}^{(\delta)}\cos (\phi_j-\phi_i)
\sin(\delta)\approx \omega_0-K\delta$, i.e.\ frequency suppression, for large values of $K$ and small values of $\delta$.
Here we used properties of the sine function and of the synchronized solution.
 
%%%%%%%%%%%%%%%%%%%%%%%%%%%%%%%%%%%%%%%%%%%%%%%%%%%%%%%%%%%%%%%%%%%
\section{Calculation of the instantaneous largest Lyapunov exponent}

In this section, we describe the algorithm that we use to calculate estimates for the instantaneous largest Lyapunov exponent $\lambda_{max}(t)$ together with the numerical integration of Eq.~(6).
Since solutions of Eq.~(6) are vectors of phases $\bm{\phi}=(\phi_1,\dots \phi_N)$, an appropriate distance function $d(\bm{\phi},\bm{\psi})$ on the $N$-torus, measuring the separation between reference orbits and perturbed orbits at each time step, is needed. 
We choose the $1$-norm distance 
\begin{equation}
d(\bm{\phi},\bm{\psi}) = \sum_{i=1}^N\left|\phi_i-\psi_i\right|\quad,
\end{equation}
since it maximizes all $p$-norm distances, thus providing an upper bound on the separation distance. 
Note that making use of the embedding into Euclidean space $\bm{\phi}\to(\sin(\bm{\phi}),\cos(\bm{\phi}))$ together with applying the $2$-norm runs into obvious problems in the renormalization step of the algorithm (see below).

An initial separation between the reference trajectory and the perturbed trajectory of $d_0 = 10^{-4}$ is chosen. 
Both trajectories are integrated for one time step $dt=0.1$ according to Eq.~(6), and the quantity $\lambda_1=\ln(d_1/d_0)/dt$ is calculated. 
$d_1$ is the separation between the trajectories after one step. 
For small initial separations $d_0$ this quantity is a good approximation for the instantaneous Lyapunov exponent after one time step~\cite{lyap}. 
Now the perturbed trajectory is renormalized in the direction of the difference vector to the reference trajectory to have a separation of $d_0$ again. 
After another integration step the quantity $\lambda_2=\ln(d_2/d_0)+\ln(d_1/d_0)/(2dt)$ is calculated and the perturbed trajectory is renormalized as before. 
Iteration yields estimates for the instantaneous largest Lyapunov exponents 
\begin{equation}
\lambda_{max}(t_n)=\lambda_n=\frac{1}{t_n-t_0}\sum_{i=1}^n\ln\left(\frac{d_i}{d_0}\right)\quad.
\end{equation}
The algorithm was tested in a discrete version with the logistic map; $d_0 = 10^{-4}$ turned out to be sufficiently small to provide satisfactory results.
\pagebreak
\end{widetext}
%%%%%%%%%%%%%%%%%%%%%%%%%%%%%%%%%%%%%%%%%%%%%%%%%%%%%%%%%%%%%%%%%%%

\begin{thebibliography}{99}
	\bibitem{osci} G.~Buzs\'{a}ki and A.~Draguhn,
	%\textit{Neuronal oscillations in cortical networks}, 
	Science \textbf{304}, 1926 (2004).
	%(5679)
	\bibitem{epilepsy} M.~Eisenstein,
	%\textit{Neurobiology: Unrestrained excitement}, 
	Nature \textbf{511}, 4 (2014).
	%(7508)
	\bibitem{schizo} D.~A.~Lewis, T.~Hashimoto, and D.~W.~Volk,
	%\textit{Cortical inhibitory neurons and schizophrenia}, 
	Nat. Rev. Neurosci. \textbf{6}, 312 (2005).
	%(4)
	\bibitem{brainweb} F.~Varela, J.~P.~Lachaux, E.~Rodriguez, and J.~Martinerie,
	%\textit{The brainweb: phase synchronization and large-scale integration}, 
	Nat. Rev. Neurosci. \textbf{2}, 229 (2001).
	%(4)
	\bibitem{eeg} C.~J.~Stam,
	%\textit{Nonlinear dynamical analysis of EEG and MEG: Review of an emerging field}, 
	J. Clin. Neurophysiol. \textbf{116}, 2266 (2005).
	\bibitem{cbn} E.~Bullmore and O.~Sporns,
	%\textit{Complex brain networks: graph theoretical analysis of structural and functional systems}, 
	Nat. Rev. Neurosci. \textbf{10}, 186 (2009).
	%(3)
	\bibitem{inhib}  J.~S.~Isaacson and M.~Scanziani,
	%\textit{How inhibition shapes cortical activity}, 
	Neuron \textbf{72}, 231 (2011).
	%(2)
	\bibitem{rsn} B.~B.~Biswal,
	%\textit{Resting state fMRI: A personal history}, 
	Neuroimage \textbf{62}, 938 (2012).
	%(2)
	\bibitem{deco} G.~Deco, V.~K.~Jirsa, and A.~R.~McIntosh,
	%\textit{Emerging concepts for the dynamical organization of resting-state activity in the brain}, 
	Nat. Rev. Neurosci. \textbf{12}, 43 (2010).
	%(1)
	\bibitem{honey} C.~J.~Honey, R.~K\"otter, M.~Breakspear, and O.~Sporns,
	%\textit{Network structure of cerebral cortex shapes functional connectivity on multiple time scales}, 
	Proc. Nat. Acad. Sci. USA \textbf{104}, 10240 (2007).
	%(24)
	\bibitem{ghosh} A.~Ghosh, Y.~Rho, A.~R.~McIntosh, R.~K\"otter, and V.~K.~Jirsa,
	%\textit{Noise during Rest Enables the Exploration of the BrainÕs Dynamic Repertoire}, 
	PLoS Comput. Biol. \textbf{4}, e1000196 (2008).
	%(10)
	\bibitem{deco2} G.~Deco, V.~Jirsa, A.~R.~McIntosh, O.~Sporns, and R.~K\"otter,
	%\textit{Key role of coupling, delay, and noise in resting brain fluctuations} 
	Proc. Nat. Acad. Sci. USA \textbf{106}, 10302 (2009).
	%(25)
	\bibitem{rs} J.~Cabral, E.~Hugues, O.~Sporns, and G.~Deco, 
	%\textit{Role of local network oscillations in resting-state functional connectivity}, 
	Neuroimage \textbf{57}, 130 (2011).
	%(1)
	\bibitem{gray} W.~Singer and C.~M.~Gray,
	%\textit{Visual feature integration and the temporal correlation hypothesis}, 
	Annu. Rev. Neurosci. \textbf{18}, 555 (1995).
	%(1)
	\bibitem{gamma} G.~Buzs\'{a}ki and X.~J.~Wang, 
	%\textit{Mechanisms of gamma oscillations}, 
	Annu. Rev. Neurosci. \textbf{35}, 203 (2012).
	\bibitem{whittington} M.~A.~Whittington, R.~D.~Traub, and J.~G.~R.~Jefferys,
	%\textit{Synchronized oscillations in interneuron networks driven by metabotropic glutamate receptor activation}, 
	Nature \textbf{373}, 612 (1995).
	%(6515)
	\bibitem{ii} X.~J.~Wang and G. Buzs\'{a}ki,
	%\textit{Gamma Oscillation by Synaptic Inhibition in a Hippocampal Interneuronal Network Model}, 
	J. Neurosci. \textbf{16}, 6402 (1996).
	%(20)
	\bibitem{brunel} N.~Brunel and X.~J.~Wang,
	%\textit{What determines the frequency of fast network oscillations with irregular neural discharges? I. Synaptic dynamics and excitation-inhibition balance}, 
	J. Neurophysiol. \textbf{90}, 415 (2003).
	%(1)
	\bibitem{geisler} C.~Geisler, N.~Brunel, and X.~J.~Wang,
	%\textit{Contributions of intrinsic membrane dynamics to fast network oscillations with irregular neuronal discharges}, 
	J. Neurophysiol. \textbf{94}, 4344 (2005).
	%(6)
	\bibitem{wc} H.~R.~Wilson and J.~D.~Cowan,
	%\textit{Excitatory and inhibitory interactions in localized populations of model neurons}, 
	Biophys. J. \textbf{12}, 1 (1972).
	%(1)
	\bibitem{mice} E.~O.~Mann and I.~Mody,
	%\textit{Control of hippocampal gamma oscillation frequency by tonic inhibition and excitation of interneurons}, 
	Nat. Neurosci. \textbf{13}, 205 (2010).
	%(2)
	\bibitem{rats} A.~V.~Medvedev,
	%\textit{Epileptiform spikes desynchronize and diminish fast (gamma) activity of the brain: an ``anti-binding'' mechanism?}, 
	Brain Res. Bull. \textbf{58}, 115 (2002).
	%(1)
	\bibitem{gaba} S.~D.~Muthukumaraswamy, R.~A.~E.~Edden, D.~K.~Jones, J.~B.~Swettenham, and K.~D.~Singh,
	%\textit{Resting GABA concentration predicts peak gamma frequency and fMRI amplitude in response to visual stimulation in humans}, 
	Proc. Nat. Acad. Sci. USA \textbf{106}, 8356 (2009).
	%(20)
	\bibitem{bk} R.~M.~Borisyuk and A.~B.~Kirillov,
	%\textit{Bifurcation analysis of a neural network model}, 
	Biol. Cybern. \textbf{66}, 319 (1992).
	%(4)
	\bibitem{izh} F.~C.~Hoppensteadt and E.~M.~Izhikevich, \textit{Weakly connected neural networks} (Springer Series Applied Mathematical Sciences, New York, 1997), Vol. 126.
	\bibitem{fairen} A.~Fairen, J.~DeFelipe, and J.~Regidor, 
	%\textit{Nonpyramidal neurons: general account}, 
	Cereb. Cortex \textbf{7}, 347 (1984).
	%(4)
	\bibitem{defelipe} J.~DeFelipe, I.~Fari\~{n}as,
	%\textit{The pyramidal neuron of the cerebral cortex: morphological and chemical characteristics of the synaptic inputs}, 
	Prog. Neurobiol. \textbf{39}, 563 (1992).
	%(6)
	\bibitem{supp} See supplemental material for details concerning the derivation of the multiplex Kuramoto model, the case $\delta=0$ and the calculation of the instantaneous largest Lyapunov exponent.
	\bibitem{dfmri} T.~Tsurugizawa, L.~Ciobanu, and D.~Le Bihan,
	%\textit{Water diffusion in brain cortex closely tracks underlying neuronal activity}, 
	Proc. Nat. Acad. Sci. USA \textbf{110}, 11636 (2013).
	%(28)
	\bibitem{winfree} A.~Winfree,
	%\textit{Biological rhythms and the behavior of populations of coupled oscillators}, 
	J. Theor. Biol. \textbf{16}, 15 (1967).
	%(1)
	\bibitem{kura} Y.~Kuramoto, \textit{Self-entrainment of a population of coupled non-linear oscillators},
	in \textit{Int. Symp. on Math. Prob. in Theor. Phys.}, ed. H. Araki (Springer Series Lecture Notes in Physics, Berlin Heidelberg, 1975), Vol. 39.
	\bibitem{kura2} Y.~Kuramoto,
	%\textit{Cooperative dynamics of oscillator community}, 
	Prog. Theor. Phys. Supp. \textbf{79}, 223 (1984).
	\bibitem{arenas} A.~Arenas, A.~D\'{i}az-Guilera, J.~Kurths, Y.~Moreno, and C.~Zhou, 
	%\textit{Synchronization in complex networks}, 
	Phys. Rep. \textbf{469}, 93 (2008).
	%(3)
	\bibitem{kalloniatis} A.~C.~Kalloniatis,
	%\textit{From incoherence to synchronicity in the network Kuramoto model}, 
	Phys. Rev. E \textbf{82}, 066202 (2010).
	%(6)
	\bibitem{llekura} G.~Miritello, A.~Pluchino, and A.~Rapisarda,
	%\textit{Central limit behavior in the Kuramoto model at the ``edge of chaos''}, 
	Physica A \textbf{388}, 4818 (2009).
	%(23)
	\bibitem{niebur} E.~Niebur, H.~G.~Schuster, and D.~M.~Kammen,
	%\textit{Collective frequencies and metastability in networks of limit-cycle oscillators with time delay}, 
	Phys. Rev. Lett. \textbf{67}, 2753 (1991).
	%(20)
	\bibitem{yeung} M.~K.~S.~Yeung and S.~H.~Strogatz,
	%\textit{Time delay in the Kuramoto model of coupled oscillators}, 
	Phys. Rev. Lett. \textbf{82}, 648 (1999).
	%(3)
	\bibitem{enzo} V.~Nicosia, M.~Valencia, M.~Chavez, A.~D\'{i}az-Guilera, and V.~Latora,
	%\textit{Remote synchronization reveals network symmetries and functional modules}, 
	Phys. Rev. Lett. \textbf{110}, 174102 (2013).
	%(17)
	\bibitem{acebron} J.~A.~Acebr\'{o}n, L.~L.~Bonilla, C.~J.~P.~Vicente, F.~Ritort, and R.~Spigler,
	%\textit{The Kuramoto model: A simple paradigm for synchronization phenomena}, 
	Rev. Mod. Phys. \textbf{77}, 137 (2005).
	%(1)
	\bibitem{howto} V.~H.~P.~Louzada, N.~A.~M. Ara\'ujo, J.~S.~Andrade, Jr., and H.~J.~Herrmann,
	%\textit{How to suppress undesired synchronization}, 
	Nat. Sci. Rep. \textbf{2} (2012).
	\bibitem{choi} M.~Y.~Choi, H.~J.~Kim, D.~Kim, and H.~Hong,
	%\textit{Synchronization in a system of globally coupled oscillators with time delay}, 
	Phys. Rev. E \textbf{61}, 371 (2000).
	%(1)
	\bibitem{brod} K.~Zilles and K.~Amunts,
	%\textit{Centenary of BrodmannÕs map Ñ conception and fate}, 
	Nat. Rev. Neurosci. \textbf{11}, 139 (2010).
	%(2)
	\bibitem{parc} D.~C.~Van Essen,  M.~F.~Glasser, D.~L.~Dierker, J.~Harwell, and T.~Coalson, 
	%\textit{Parcellations and hemispheric asymmetries of human cerebral cortex analyzed on surface-based atlases}, 
	Cereb. Cortex \textbf{22}, 2241 (2011).
	%(10)
	\bibitem{lyap} S.~H.~Strogatz, \textit{Nonlinear dynamics and chaos: with applications to physics, biology, chemistry, and engineering} 
	(Perseus Books Group, New York, 1994).
\end{thebibliography}
\end{document}